\documentclass[12pt,a4paper]{article}%
\usepackage{amsfonts}
\usepackage{graphicx}
\usepackage{amsmath}
\usepackage{amssymb}%
\providecommand{\U}[1]{\protect\rule{.1in}{.1in}}

\begin{document}

\title{\textbf{Testing for change in mean of heteroskedastic time series}%
}
\author{Mohamed BOUTAHAR \thanks{%
Departement of Mathematics, case 901. Luminy Faculty of Sciences.\ 163 Av.
de Luminy 13288 Marseille\ Cedex 9.\ GREQAM, 2 rue de la charit\'{e} 13002.
\ e-mail: boutahar@univmed.fr.} \\
University of M\'{e}diterran\'{e}e,\ Marseille, France}
\maketitle

\bigskip

\textbf{Abstract. }

In this paper we consider a Lagrange Multiplier-type test (LM) to detect
change in the mean of time series with heteroskedasticity of unknown form.
We derive the limiting distribution under the null, and prove the
consistency of the test against the alternative of either an abrupt or
smooth changes in the mean. We perform also some Monte Carlo simulations to
analyze the size distortion and the power of the proposed test. We conclude
that for moderate sample size, the test has a good performance. We finally
carry out an empirical application using the daily closing level of the S\&P
500 stock index, in order to illustrate the usefulness of the proposed test.

\textit{AMS classifications codes:} 62G10, 62G20, 60F17, 62M10


\section{Introduction}

In the statistic literature there is a vast amount of works on detecting
changes in mean of a given time series. In a more general context of linear
regression model, Chow (1960) considered tests for structural change for a
known single break date. The researches headed for the modelling where this
break date is treated as an unknown variable. Quandt (1960) extends the Chow
test and proposes taking the largest Chow statistic over all possible break
dates. In the same context, the most important contributions are those of
Andrews (1993) and Andrews and Ploberger (1994). Sen and Srivastava (1975a,
1975b), Hawkins (1977), Worsley (1979), Srivastava and Worsley (1986) and
James et al. (1987) consider tests for mean shifts of normal sequence of
variables. The multiple structural changes case receives an increasing
attention. For instance, Yao (1988), Yin (1988) and Yao and Au (1989) study
the estimation of the number of mean shifts of variables sequence using the
Bayesian information criterion. Liu et al. (1997) consider multiple changes
in a linear model estimated by least squares and estimate the number of
changes using a modified Schwarz' criterion. Bai and Perron (1998) consider
the estimation of multiple structural shifts in a linear model estimated by
least squares; Qu and Perron (2007) extend Bai and Perron's (1998) results
to a multivariate regression. In all these papers, a Wald, Lagrange
Multiplier (LM) or/and Likelihood-ratio (LR)-Like tests have been
considered. Recall that the Wald test is based on the unrestricted model,
the LR test needs the restricted and unrestricted model, while the LM test
is based exclusively on the restricted model.

Concerning only the change in mean, all authors cited above assume that
under the alternative hypothesis, the mean $\mu _{t}$ is a step function
i.e. the observations $(y_{t}),$ $1\leq t\leq n,$ satisfy
\begin{eqnarray*}
y_{t} &=&\mu _{t}+\text{ }\varepsilon _{t}, \\
\mu _{t} &=&\mu _{(j)\text{ }}\text{if }t=n_{j-1}+1,...,n_{j},n_{j}=[\lambda
_{j}n],0<\lambda _{1}<...<\lambda _{m}<1,
\end{eqnarray*}%
where $(\varepsilon _{t})$ is such that E$(\varepsilon _{t})=0$ and $[x]$\
is the integer part of $x$. If the mean $\mu _{t}$ is time varying with
unknown form, then the Wald and LR tests can't be applied. Only the LM test
can be used since no specification of alternative hypothesis is needed to
build a statistic. Recently, Gombay (2008) used an LM-type test for
detecting change in the autoregressive model. However, he assumed that the
errors $\varepsilon _{t}$ are homoskedastic i.e var($\varepsilon
_{t})=\sigma ^{2\text{ }}$ for all $t$. In this paper we consider the
heteroskedastic time series,
\begin{equation}
y_{t}=\mu _{t}+\sigma _{t}\varepsilon _{t},  \label{yt}
\end{equation}%
where the errors are Gaussian white noise $\varepsilon _{t}\thicksim N(0,1)$%
, $(\sigma _{t})$ is a deterministic sequence with unknown form. The null
and the alternative hypotheses are as follows:
\begin{equation}
\left\{
\begin{tabular}{l}
$H_{0}:\mu _{t}=\mu $ for all $t\geq 1$ \\
against \\
$H_{1}:$ There exist $t\neq s$ such that $\mu _{t}\neq \mu _{s}$%
\end{tabular}%
\ \ \right. ,  \label{h01}
\end{equation}%
under the alternative hypothesis the mean $\mu _{t}$ can be time varying
with unknown form. The model (\ref{yt}) is useful in many areas. In
financial modelling, much research has been devoted to the study of long-run
behavior of returns of speculative asset. A common finding in much of the
empirical literature is that the returns are not serially correlated which
is in agreement with the efficient market hypotheses, see Ding et al.
(1993). However, the absolute returns which, is a proxy of the instantaneous
standard deviation, has significant positive autocorrelations with a
possible breaks in the mean and in the unconditional variance.\ For
instance, Starica and Granger (2005) show that an appropriate model to
describe the dynamic of the logarithm of the absolute returns of the S\&P
500 index is given by (\ref{yt}) where\ $\mu _{t}$ and $\sigma _{t}$ are
step functions i.e.

\begin{eqnarray}
\mu _{t} &=&\mu _{(j)\text{ }}\text{if }t=n_{j-1}+1,...,n_{j},n_{j}=[\lambda
_{j}n],0<\lambda _{1}<...<\lambda _{m_{1}}<1,  \label{m} \\
\sigma _{t} &=&\sigma _{(j)\text{ }}\text{if }t=t_{j-1}+1,...,t_{j},t_{j}=[%
\tau _{j}n],0<\tau _{1}<...<\tau _{m_{2}}<1,  \label{s}
\end{eqnarray}%
for some integers $m_{1}$ and $m_{2}$. They also show that model (\ref{yt}),
(\ref{m}) and (\ref{s}) gives forecasts superior to those based on a
stationary GARCH(1,1) model.

\bigskip One can also consider a more general model than (\ref{yt}), (\ref{m}%
) and (\ref{s}), where breaks can be abrupt and/or smooth. A model with $%
(m+1)$ regimes for the unconditional standard deviation can be defined by
\begin{equation}
\sigma _{t}=\sum_{j=1}^{m}\sigma _{(j)\text{ }}\left( 1-F_{j}\left( \frac{%
t/n-\tau _{2j-1}}{s_{j}}\right) \right) +\sigma _{(j+1)\text{ }}F_{j}\left(
\frac{t/n-\tau _{2j-1}}{s_{j}}\right) ,  \label{sd2}
\end{equation}%
where $\tau _{0}=0<\tau _{1}<...<\tau _{2m}<\tau _{2m+1}=1,$ $F_{j}$ is the
transition function from regime $j$ to regime $(j+1)$, assumed to be
continuous from $%
\mathbb{R}
$ onto $[0,1].$ The scale $s_{j}>0$\ indicates how rapidly the transition
from regime $j$ to regime $(j+1)$, a small $s_{j}$ yields an abrupt change,

As in Gombay (2008) we use an LM-type test for detecting change in mean. The
test statistic is based on the normalized score vector evaluated under the
null $H_{0}^{\prime }:\mu _{t}=\mu $ and $\sigma _{t}=\sigma $ for all $%
t\geq 1.$ If $y_{t}=\mu +\sigma \varepsilon _{t}$ then the log-likelihood of
the sample is given by

\begin{equation*}
L(n,\mu ,\sigma ^{2})=-\frac{n}{2}\log \sigma ^{2}-\frac{n}{2}\log 2\pi -%
\frac{1}{2\sigma ^{2}}\sum_{t=1}^{n}(y_{t}-\mu )^{2}.
\end{equation*}%
Hence, the score vector is
\begin{equation*}
S_{n}(\mu ,\sigma ^{2})=\left(
\begin{array}{l}
\frac{1}{\sigma ^{2}}\sum_{t=1}^{n}(y_{t}-\mu ) \\
-\frac{n}{2\sigma ^{2}}+\ \frac{1}{2\sigma ^{4}}\sum_{t=1}^{n}(y_{t}-\mu
)^{2}%
\end{array}%
\right)
\end{equation*}%
and the information matrix is $I_{n}(\mu ,\sigma ^{2})=nI/\sigma ^{2},$ $I$
is the identity matrix. Therefore a test statistic for testing change in
mean is based on the first component of the vector $I_{n}^{-1/2}(\widehat{%
\mu },\widehat{\sigma }^{2})S_{[n\tau ]}(\widehat{\mu },\widehat{\sigma }%
^{2}),$ where $\widehat{\mu }=\sum_{t=1}^{n}y_{t}/n$ and $\widehat{\sigma }%
^{2}=\sum_{t=1}^{n}(y_{t}-\widehat{\mu })^{2}/n$ are the maximum likelihood
estimators of $\mu $ and $\sigma ^{2},$ given by
\begin{equation*}
B_{n}(\tau )=\frac{1}{\sqrt{n}\widehat{\sigma }}\sum_{t=1}^{[n\tau ]}(y_{t}-%
\widehat{\mu }).
\end{equation*}%
The test statistic we consider is%
\begin{equation*}
\mathcal{B}_{n}=\sup_{\tau \in \lbrack 0,1]}\left\vert B_{n}(\tau
)\right\vert .
\end{equation*}

\section{Limiting distribution of $\mathcal{B}_{n}$ under the null}

\textbf{Theorem 1}. \textit{Assume that }$(y_{t})$\textit{\ satisfies the
model (\ref{yt}) with standard Gaussian white noise errors }$(\varepsilon
_{t})$ \textit{and a bounded deterministic sequence }$(\sigma _{t})$ \textit{%
satisfying }%
\begin{equation}
\frac{1}{n}\sum_{t=1}^{n}\sigma _{t}^{2}\rightarrow \overline{\sigma }%
_{2}^{2}\text{ \textit{as} }n\rightarrow \infty ,  \label{sig}
\end{equation}%
\textit{Then, under }$H_{0}$\textit{\ we have}
\begin{equation}
\mathcal{B}_{n}\text{ \ \ }\underrightarrow{\text{ \ \ \ }\text{ \ \ }}\text{ \ }\mathcal{B}_{\infty }=\sup_{\tau \in
\lbrack 0,1]}\left\vert B(\tau )\right\vert  \label{cv2}
\end{equation}%
$\underrightarrow{\text{ \ \ }\mathcal{L}\text{ \ }}\ $\textit{\ denotes the
convergence in distribution and }$B(\tau )$\textit{\ is a Brownian Bridge.}

Remark. The condition (\ref{sig}) is a classical ergodic assumption and
holds in many situations. For example if $(\sigma _{t})$ is given by (\ref{s}%
), then (\ref{sig}) is satisfied with $\overline{\sigma }_{2}^{2}=%
\sum_{j=1}^{m_{2}+1}(\tau _{j}-\tau _{j-1})\sigma _{(j)}^{2},$ $\tau _{0}=0,$
$\tau _{m_{2}+\text{ }1}=1,$ and if $(\sigma _{t})$\ is given by (\ref{sd2})
then
\begin{equation*}
\overline{\sigma }_{2}^{2}=\int_{0}^{1}\left\{ \sum_{j=1}^{m}\sigma _{(j)%
\text{ }}\left( 1-F_{j}\left( \frac{x-\tau _{2j-1}}{s_{j}}\right) \right)
+\sigma _{(j+1)\text{ }}F_{j}\left( \frac{x-\tau _{2j-1}}{s_{j}}\right)
\right\} ^{2}dx.
\end{equation*}

\bigskip The proof of Theorem 1 is given in the Appendix.

\section{ Consistency of $\mathcal{B}_{n}$}

\subsection{Consistency of $\mathcal{B}_{n}$ against abrupt changes}

Without loss of generality we assume that under the alternative hypothesis
there is a single break date, i.e. $(y_{t})$\textit{\ }is given by (\ref{yt}%
) where

\begin{equation}
\mathit{\mu }_{t}\mathit{=}\left\{
\begin{tabular}{l}
$\mu _{(1)}$ if $1\leq t\leq \lbrack n\tau _{1}]$ \\
$\mu _{(2)}$ if $[n\tau _{1}]+1\leq t\leq n$%
\end{tabular}%
\ \right. \text{for some }\mathit{\tau }_{1}\mathit{\in (0,1).}  \label{muta}
\end{equation}

\textbf{Theorem 2}. \textit{Assume that }$(y_{t})$\textit{\ satisfies the
model (\ref{yt}) with standard Gaussian white noise errors }$(\varepsilon
_{t})$\textit{\ and a bounded deterministic sequence }$(\sigma _{t})$
\textit{satisfying (\ref{sig}). If under }$H_{1}$\textit{\ the mean }$\mu
_{t}$\textit{\ follows the dynamic (\ref{muta}) then\ }%
\begin{equation}
\mathcal{B}_{n}\text{ \ }\underrightarrow{\text{ \ \ \ }P\text{ \ \ }}\text{
\ }+\infty  \label{bh1}
\end{equation}%
\textit{\ where }$\underrightarrow{\text{ \ \ \ }P\text{ \ \ }}$\textit{\ \
denotes the convergence in probability.}

The proof of Theorem 2 is given in the Appendix.

\textbf{Remark 1. }The result of Theorem 2 remains valid if under the
alternative hypothesis there are multiples breaks in the mean.

\subsection{$\protect\bigskip$Consistency of $\mathcal{B}_{n}$ against
smooth changes}

In economics and finance, multiple regimes modelling becomes more and more
important in order to take into account phenomena characterized, for
instance, by recession or expansion periods, or high or low volatility
periods. Consequently, it's more realistic to assume that the break in the
mean doesn't happen suddenly but the transition from one regime to another
is continuous with slowly variation. A well known dynamic is the smooth
transition autoregressive (STAR) specification, see Terasv\"{\i}rta \cite{te}%
, in which the mean $\mu _{t}$ is a time varying with respect to the
following

\begin{equation}
\mu _{t}=\mu _{(1)}+(\mu _{(2)}-\mu _{(1)})F(t/n,\tau _{1},\gamma ),1\leq
t\leq n,  \label{muts}
\end{equation}%
where $F(x,\tau _{1},\gamma )$ is a the smooth transition function assumed
to be continuous from $[0,1]$ onto $[0,1]$. The parameters $\mu _{(1)}$ and $%
\mu _{(2)}$ are the values of the mean in the two extreme regimes, that is
when $F$ $\rightarrow 0$ and $F$ $\rightarrow 1$. The slope parameter $%
\gamma $ indicates how rapidly the transition between two extreme regimes
is. The parameter $c$ is the location parameter.

Two choices for the function $F$ are frequently evoked, the logistic
function given by%
\begin{equation}
F_{L}(x,\tau _{1},\gamma )=\left[ 1+\exp (-\gamma (x-\tau _{1})\right] ^{-1}
\label{logistic}
\end{equation}%
and the exponential one

\begin{equation}
F_{e}(x,\tau _{1},\gamma )=1-\exp (-\gamma (x-\tau _{1})^{2}).
\end{equation}%
For example, for the logistic function with $\gamma >0$, the extreme regimes
are obtained as follows

\textbullet\ if $x\rightarrow 0$ and $\gamma $ large we have $F\rightarrow 0
$ and thus $\mu _{t}=\mu _{(1)}$,

\textbullet\ if $x\rightarrow 1$ and $\gamma $ large we have $F\rightarrow 1$
and thus $\mu _{t}=\mu _{(2)}$.

\textbf{Theorem 3}. \textit{Assume that }$(y_{t})$\textit{\ satisfies (\ref%
{yt}) with standard Gaussian white noise errors }$(\varepsilon _{t})$\textit{%
\ and a bounded deterministic sequence }$(\sigma _{t})$ \textit{satisfying (%
\ref{sig}). If under }$H_{1}$\textit{\ the mean }$\mu _{t}$\textit{\ follows
the dynamic (\ref{muts}) then }%
\begin{equation}
\mathcal{B}_{n}\text{ \ }\underrightarrow{\text{ \ \ \ }P\text{ \ \ }}\text{
\ }+\infty .  \label{bh2}
\end{equation}%
The proof of Theorem 3 is given in the Appendix.

\section{Finite sample performance}

All sequences are driven by a Gaussian white noise $\varepsilon
_{t}\thicksim N(0,1).$ Simulations were performed using the software $R$
\cite{r}. We carry out an experiment of $1000$ samples for nine series and
we use four different sample sizes, $n=30$, $\ n=100,$ $n=500$ and $n=1000$.

In the model (\ref{yt}) we consider three dynamics for the mean $\mu _{t}$
and the variance~$\sigma _{t}^{2}:$

$\bullet $ \textbf{Dynamics of the mean:}

\begin{equation}
\mu _{t}=1\text{ for all }t\geq 0,  \label{s1}
\end{equation}

\bigskip

\begin{equation}
\mu _{t}=\left\{
\begin{tabular}{l}
$\mu _{(1)}$ if $1\leq t\leq \lbrack n\tau _{1}]$ \\
$\mu _{(2)}$ if $[n\tau _{1}]+1\leq t\leq n$%
\end{tabular}%
\ \right. ,  \label{s2}
\end{equation}

\bigskip

\begin{equation}
\mu _{t}=\mu _{(1)}+(\mu _{(2)}-\mu _{(1)})F(t/n,\tau _{1},\gamma ),
\label{s3}
\end{equation}%
we choose $\tau _{1}=0.5$ (one break in the middle of the sample), $\mu
_{(1)}=1,\mu _{(2)}=2,$ $F$ is the logistic function given by (\ref{logistic}%
) and $\gamma =20$.

$\bullet $ \textbf{Dynamics of the variance}

\begin{equation}
\sigma _{t}=1\text{ for all }t\geq 0,  \label{s4}
\end{equation}

\bigskip

\begin{equation}
\sigma _{t}=\left\{
\begin{tabular}{l}
$\sigma _{(1)}$ if $1\leq t\leq \lbrack n\tau _{2}]$ \\
$\sigma _{(2)}$ if $[n\tau _{2}]+1\leq t\leq n$%
\end{tabular}%
\ \right.  \label{s5}
\end{equation}

\bigskip

\begin{equation}
\sigma _{t}=\sigma _{(1)}+(\sigma _{(2)}-\sigma _{(1)})F(t/n,\tau
_{2},\gamma ),  \label{s66}
\end{equation}

$\bigskip $ we choose $\tau _{2}=2/3$, $\sigma _{(1)}=0.5,\sigma _{(2)}=1.5,$
$F$ is the logistic function given by (\ref{logistic}) and $\gamma =20.$

To study the size of the test we simulate the following three series:

\textit{Series 1}: $\mu _{t}$ is given by (\ref{s1}) and $\sigma _{t}$ is
given by (\ref{s4}), no break in the mean and in the variance.

\textit{Series 2}: $\mu _{t}$ is given by (\ref{s1}) and $\sigma _{t}$ is
given by (\ref{s5}), no break in the mean and one abrupt break in the
variance.

\textit{Series 3}: $\mu _{t}$ is given by (\ref{s1}) and $\sigma _{t}$ is
given by (\ref{s66}), no break in the mean and one smooth break in the
variance.

To study the power of the test we simulate the following six series:

$\bullet $ One abrupt change on the mean:

\textit{Series 4}: $\mu _{t}$ is given by (\ref{s2}) and $\sigma _{t}$ is
given by (\ref{s4}), one abrupt break in the mean and no break in the
variance.

\textit{Series 5}: $\mu _{t}$ is given by (\ref{s2}) and $\sigma _{t}$ is
given by (\ref{s5}), one abrupt break in the mean and one abrupt break in
the variance.

\textit{Series 6}: $\mu _{t}$ is given by (\ref{s2}) and $\sigma _{t}$ is
given by (\ref{s66}), one abrupt break in the mean and one smooth break in
the variance.

$\bullet $ A smooth change in the mean:

\textit{Series 7}: $\mu _{t}$ is given by (\ref{s3}) and $\sigma _{t}$ is
given by (\ref{s4}), one smooth break in the mean and no break in the
variance.

\textit{Series 8}: $\mu _{t}$ is given by (\ref{s3}) and $\sigma _{t}$ is
given by (\ref{s5}), one smooth break in the mean and one abrupt break in
the variance.

\textit{Series 9}: $\mu _{t}$ is given by (\ref{s3}) and $\sigma _{t}$ is
given by (\ref{s66}), one smooth break in the mean and one smooth break in
the variance.

\begin{center}
$\bigskip $

$%
\begin{array}{l}
\text{\textbf{Table 1}{\small .\ Empirical test sizes (in \%)}} \\
\begin{tabular}{|l|l|l|l|l|}
\hline
$%
\begin{array}{llll}
&  &  &  \\
\text{ \ \ \ \ \ } &  &  & n \\
& \text{ \ \ \ \ \ \ \ \ \ }\alpha &  &
\end{array}%
$ & $n=30$ & $n=100$ & $n=500$ & $n=1000$ \\ \hline
$%
\begin{array}{lll}
&  & 1\% \\
Series\text{ }1 &  & 5\% \\
&  & 10\%%
\end{array}%
$ & $%
\begin{array}{l}
0.2\  \\
2.9 \\
5.1%
\end{array}%
$ & $%
\begin{array}{l}
0.4 \\
3.3 \\
7.9%
\end{array}%
$ & $%
\begin{array}{l}
0.7\text{ } \\
3.8 \\
8.2%
\end{array}%
$ & $%
\begin{array}{l}
\text{ }0.5 \\
\text{ }4.1 \\
\text{ }8.4%
\end{array}%
$ \\ \hline
$%
\begin{array}{lll}
&  & 1\% \\
Series\text{ }2 &  & 5\% \\
&  & 10\%%
\end{array}%
$ & $%
\begin{array}{l}
0.3 \\
3.4 \\
7.1%
\end{array}%
$ & $%
\begin{array}{l}
0.9 \\
5.1 \\
10.6%
\end{array}%
$ & $%
\begin{array}{l}
1.3\  \\
6.2 \\
11.7%
\end{array}%
$ & $%
\begin{array}{l}
1.3 \\
6.3 \\
12.4%
\end{array}%
$ \\ \hline
$%
\begin{array}{lll}
&  & 1\% \\
Series\text{ }3 &  & 5\% \\
&  & 10\%%
\end{array}%
$ & $%
\begin{array}{l}
0.5\  \\
4.3 \\
7.9%
\end{array}%
$ & $%
\begin{array}{l}
\text{ }0.9 \\
\text{ }4.9\  \\
\text{ }10.1%
\end{array}%
$ & $%
\begin{array}{l}
1.1\text{ } \\
6.4 \\
12.7%
\end{array}%
$ & $%
\begin{array}{l}
1.1 \\
6.3 \\
12.4%
\end{array}%
$ \\ \hline
\end{tabular}%
\end{array}%
$
\end{center}

{\small Note: Table 1 contains rejection frequencies of the null hypothesis
of no change in the mean. Rejection frequencies are based on 1000
replications generated from the \textit{Series} 1-3} {\small where the
nominal significance levels are 1\% , 5\% and 10\%, the sample sizes are }$%
{\small n=30,n=100,n=500}$ {\small and }${\small n=1000.}$

Table 1 indicates that the test is a somewhat conservative (the empirical
size is lesser than the nominal one) when the time series is homoskedastic (%
\textit{Series} 1) \ and overrejects the null (the empirical size is greater
than the nominal one) if the time series is heteroskedastic (\textit{Series}
2 and 3).

\begin{center}
$%
\begin{array}{l}
\text{\textbf{Table 2}{\small .\ Empirical test powers (in \%)}} \\
\begin{tabular}{|l|l|l|l|l|}
\hline
$%
\begin{array}{llll}
&  &  &  \\
\text{ \ \ \ \ \ } &  &  & n \\
& \text{ \ \ \ \ \ \ \ \ \ }\alpha &  &
\end{array}%
$ & $n=30$ & $n=100$ & $n=500$ & $n=1000$ \\ \hline
$%
\begin{array}{lll}
&  & 1\% \\
Series\text{ }4 &  & 5\% \\
&  & 10\%%
\end{array}%
$ & $%
\begin{array}{l}
\ 18.3\  \\
\ 47.3\  \\
\ 61.9%
\end{array}%
$ & $%
\begin{array}{l}
95.9 \\
98.8 \\
99.4%
\end{array}%
$ & $%
\begin{array}{l}
100 \\
100 \\
100%
\end{array}%
$ & $%
\begin{array}{l}
100 \\
100 \\
100%
\end{array}%
$ \\ \hline
$%
\begin{array}{lll}
&  & 1\% \\
Series\text{ }5 &  & 5\% \\
&  & 10\%%
\end{array}%
$ & $%
\begin{array}{l}
\ 10.6 \\
\ 33.9 \\
\ 48.5%
\end{array}%
$ & $%
\begin{array}{l}
85.0 \\
95.4 \\
97.7%
\end{array}%
$ & $%
\begin{array}{l}
100\text{ } \\
100 \\
100%
\end{array}%
$ & $%
\begin{array}{l}
100 \\
100 \\
100%
\end{array}%
$ \\ \hline
$%
\begin{array}{lll}
&  & 1\% \\
Series\text{ }6 &  & 5\% \\
&  & 10\%%
\end{array}%
$ & $%
\begin{array}{l}
\ 14.2 \\
\ 34.5 \\
\ 48.7%
\end{array}%
$ & $%
\begin{array}{l}
84.8 \\
94.8 \\
98.0%
\end{array}%
$ & $%
\begin{array}{l}
100\  \\
100 \\
100%
\end{array}%
$ & $%
\begin{array}{l}
100 \\
100 \\
100%
\end{array}%
$ \\ \hline
$%
\begin{array}{lll}
&  & 1\% \\
Series\text{ }7 &  & 5\% \\
&  & 10\%%
\end{array}%
$ & $%
\begin{array}{l}
\ 17.1 \\
\ 46.6 \\
\ 58.4%
\end{array}%
$ & $%
\begin{array}{l}
92.9 \\
98.4 \\
99.3%
\end{array}%
$ & $%
\begin{array}{l}
100 \\
100 \\
100%
\end{array}%
$ & $%
\begin{array}{l}
100 \\
100 \\
100%
\end{array}%
$ \\ \hline
$%
\begin{array}{lll}
&  & 1\% \\
Series\text{ }8 &  & 5\% \\
&  & 10\%%
\end{array}%
$ & $%
\begin{array}{l}
\ 12.6 \\
\ 36.0 \\
\ 52.1%
\end{array}%
$ & $%
\begin{array}{l}
79.8 \\
93.1 \\
96.5%
\end{array}%
$ & $%
\begin{array}{l}
100 \\
100 \\
100%
\end{array}%
$ & $%
\begin{array}{l}
100 \\
100 \\
100%
\end{array}%
$ \\ \hline
$%
\begin{array}{lll}
&  & 1\% \\
Series\text{ }9 &  & 5\% \\
&  & 10\%%
\end{array}%
$ & $%
\begin{array}{l}
14.0 \\
35.9 \\
50.8%
\end{array}%
$ & $%
\begin{array}{l}
74.8\  \\
92.0\ \text{ } \\
\text{ }95.5%
\end{array}%
$ & $%
\begin{array}{l}
100 \\
100 \\
100%
\end{array}%
$ & $%
\begin{array}{l}
100 \\
100 \\
100%
\end{array}%
$ \\ \hline
\end{tabular}%
\end{array}%
$
\end{center}

\bigskip {\small Note: Table 2 contains rejection frequencies of the null
hypothesis of no change in the mean. Rejection frequencies are based on 1000
replications generated from the \textit{Series} 4-9} {\small where the
nominal significance levels are 1\% , 5\% and 10\%, the sample sizes are }$%
{\small n=30,n=100,n=500}$ {\small and }${\small n=1000.}$

From Table 2, we observe that, except for the small sample size n=30, the
test has a good power either for homoskedastic time series (\textit{Series}
4 and 7) or heteroskedastic time series (\textit{Series} 5,6,8 and 9).
Rejection frequencies of the null in abrupt change (\textit{Series} 4, 5 and
6) are a somewhat greater than the ones corresponding to a smooth change (%
\textit{Series} 7,8 and 9).

\section{Application to the Stock index data}

We consider the daily returns of S\&P 500 index, $r_{t}=$ $\log P_{t}-\log
P_{t-1}$, where $P_{t}$ is the daily closing level of the index between
January 3, 1950 and November 17, 2008. We test changes in the
mean of the returns $r_{t}$ and the absolute returns $y_{t}=\left\vert
r_{t}\right\vert .$

\bigskip For the time series $(y_{t}),1\leq t\leq n,$ the test statistic is
given by

\begin{equation*}
\mathcal{B}_{n}=\frac{1}{\sqrt{n}\widehat{\sigma }}\max_{1\leq k\leq
n}\left\vert \sum_{t=1}^{k\ }y_{t}-k\widehat{\mu }\right\vert ,
\end{equation*}%
where $\widehat{\sigma }^{2}=\sum_{t=1}^{n}(y_{t}-\widehat{\mu })^{2}/n$ , $%
\widehat{\mu }=\sum_{t=1}^{n\ }y_{t}/n$ and the corresponding \emph{p-value} given
by $p-value=1-F_{B_{\infty }}(\mathcal{B}_{n})$. The cumulative distribution
function of $B_{\infty }$ is given by (see Billingsley (1968))
\begin{equation*}
F_{B_{\infty }}(z)=1+2\sum_{k=1}^{\infty }(-1)^{k}\exp \{-2k^{2}z^{2}\}.
\end{equation*}

Although the distribution function $F_{B_{\infty }}$ involves an infinite
sum, the series is extremely rapidly converging. Usually a few terms suffice
for very high accuracy. For example, (see Massey (1952)) the $90\%,95\%,$
and $99\%$ quantiles are $1.225,1.359$ and $1.628$ respectively.\ Note that
the quantiles are reached with a high accuracy using only $2$ terms i.e. $$1+2\sum_{k=1}^{2}(-1)^{k}\exp
\{-2k^{2}(1.225)^{2}\}=0.9005625,$$  $$1+2\sum_{k=1}^{2}(-1)^{k}\exp
\{-2k^{2}(1.359)^{2}\}=0.9502443$$ and $$1+2\sum_{k=1}^{2}(-1)^{k}\exp
\{-2k^{2}(1.628)^{2}\}=0.9900245$$

\bigskip
Applying our test to $r_{t}$ yields $%
p-value=0.291$ and hence the null hypothesis of no change in the mean is not
rejected.

To check if to the time series \ $y_{t}=\left\vert r_{t}\right\vert $is  affected by breaks in the mean, we apply our test
to $y_{t}$ to detect change in the mean. We obtain $p-value=0$ for $y_{t}$ ,
which strongly supports change in the mean of the absolute returns of S\&P
500 index between January 4,1950 and November 17, 2008

\textbf{Appendix. Proofs}

To prove Theorem\ 1 we will establish first a functional central limit
theorem for heteroskedastic time series. Such theorem is independent of
interest. Let $D=D[0,1]$\ be the space of random functions that are
right-continuous and have left limits, endowed with the Skorohod topology.
The weak convergence of a sequence of random elements $X_{n}$ in $D$ to a
random element $X$ in $D$ will be denoted by $X_{n}\Longrightarrow X.$

Consider a standard Gaussian white noise $(\varepsilon _{t})$, i.e. $%
E(\varepsilon _{t})=0$ and var$(\varepsilon _{t})=1.$ Let $(\sigma _{t})$
satisfying (\ref{sig}) and
\begin{equation}
W_{n}(\tau )=\frac{1}{\overline{\sigma }_{2}\sqrt{n}}\sum_{t=1}^{[n%
\tau ]}\sigma _{t}\varepsilon _{t},\text{ \ \ \ }\tau \in \lbrack 0,1].
\label{xy}
\end{equation}%
Many Functional central limit theorems were established for covariance
stationary time series, see Boutahar (2008) and the references therein. Note
that the process $(\sigma _{t}\varepsilon _{t})$ is not covariance
stationary and\ hence Davydov's (1970) results can't be applied to obtain
the weak convergence of $W_{n}$ in the Skorohod space.

There are two sufficient conditions to have $W_{n}\Longrightarrow W$ (see
Billingsley (1968):

i) the finite-dimensional distributions of $W_{n}$\ converge to the
finite-dimensional distributions of $W,$

ii) $W_{n}$\ is tight.

\textbf{Theorem\ A1}. \textit{Assume that }$(\varepsilon _{t})$\textit{\ is
a standard Gaussian white noise and }$(\sigma _{t})$\textit{\ satisfying (%
\ref{sig}). \ Then }%
\begin{equation}
W_{n\text{ \ }}\Longrightarrow \ \ W  \label{wn}
\end{equation}%
\textit{where }$W$\textit{\ is standard Brownian motion.}

\textbf{Proof}. To prove that the finite-dimensional distributions of $W_{n}$%
\ converge to those of $W$ it is sufficient to show that for all integer $r$
$\geq 1,$\ for all\ $0\leq \tau _{1}<...<\tau _{r}\leq 1$\ and\ for \ all $%
(\alpha _{1},...,\alpha _{r})^{\prime }$\ $\in
\mathbb{R}
^{r},$\
\begin{equation}
Z_{n}=\sum_{i=1}^{r}\alpha _{i}W_{n}(\tau _{i}){\large \ }\text{\ \
\ \ }\underrightarrow{\text{ \ \ \ }\text{ \ \ }}\text{
\ \ }Z=\text{\ }\sum_{i=1}^{r}\alpha _{i}W(\tau _{i}).\text{\ }%
{\large \ }\text{ }  \label{con1}
\end{equation}%
Since $Z_{n}$ is Gaussian with zero mean, it is sufficient to prove that%
\begin{equation}
\text{var}\left( Z_{n}\right) \text{\ }\rightarrow \text{ var}%
(Z)=\sum_{1\leq i,j\leq r}\alpha _{i}\ \alpha _{j}\min (\tau
_{i},\tau _{j}){\large .}  \label{zn}
\end{equation}%
For all $(\tau _{i},\tau _{j})$%
\begin{align*}
\text{cov}(W_{n}(\tau _{i}),W_{n}(\tau _{j}))& =\text{var}(W_{n}(\min (\tau
_{i},\tau _{j}) \\
& =\frac{1}{n\overline{\sigma }_{2}^{2}}\sum_{t=1}^{[n\min (\tau _{i},\tau
_{j})]}\sigma _{t}^{2} \\
& \rightarrow \min (\tau _{i},\tau _{j})\text{ as }n\rightarrow \infty ,
\end{align*}%
since var$\left( Z_{n}\right) =\sum_{1\leq i,j\leq r}\alpha _{i}\ \alpha
_{j}$cov$(W_{n}(\tau _{i}),W_{n}(\tau _{j}))$, the desired conclusion (\ref%
{zn}) holds.

To prove the tightness of\ $W_{n}$\ it suffices to show the following
inequality [Billingsley (1968), Theorem 15.6]
\begin{equation}
E\left( \left\vert W_{n}(\tau )-W_{n}(\tau _{1})\right\vert ^{\gamma
}\left\vert W_{n}(\tau _{2})-W_{n}(\tau )\right\vert ^{\gamma }\right) \leq
(F(\tau _{2})-F(\tau _{1}))^{\alpha }  \label{bil}
\end{equation}%
for some $\gamma \geq 0$, $\alpha >1,$\ and $F$\ is a nondecreasing
continuous function on [0,1], where $0<\tau _{1}<\tau <\tau _{2}<1.$

We have
\begin{eqnarray*}
E\left( \left\vert W_{n}(\tau )-W_{n}(\tau _{1})\right\vert ^{2}\left\vert
W_{n}(\tau _{2})-W_{n}(\tau )\right\vert ^{2}\right) & =& \frac{1}{n\overline{%
\sigma }_{2}^{2}}\left( \sum_{t=[n\tau _{1}]+1}^{[n\tau ]}\sigma
_{t}^{2}\right) \left( \sum_{t=[n\tau ]+1}^{[n\tau _{2}]}\sigma
_{t}^{2}\right) \\
& \leq & C(\tau -\tau _{1})(\tau _{2}-\tau ) \\
& & \text{ for some constant }C>0 \\
& \leq C&  (\tau _{2}-\tau _{1})^{2}/2.
\end{eqnarray*}
Consequently (\ref{bil}) holds with $\gamma =\alpha =2$ and $F(t)=\sqrt{C/2}$
$t$. \ \

\bigskip\ \textbf{A1. Proof of Theorem 1}

We have
\begin{align*}
B_{n}(\tau )& =\frac{1}{\sqrt{n}\widehat{\sigma }}\sum_{t=1}^{[n\tau
]}(y_{t}-\widehat{\mu }) \\
& =\left( \frac{\overline{\sigma }_{2}}{\widehat{\sigma }}\right) \frac{1}{%
\sqrt{n}\overline{\sigma }_{2}}\sum_{t=1}^{[n\tau ]}\left\{ (y_{t}-\mu
)+(\mu -\widehat{\mu })\right\} \\
& =\left( \frac{\overline{\sigma }_{2}}{\widehat{\sigma }}\right) \left\{
W_{n}(\tau )-\frac{[n\tau ]}{n}W_{n}(1)\right\} .
\end{align*}%
By using (\ref{wn}) it follows that
\begin{equation*}
\left( \frac{\widehat{\sigma }}{\overline{\sigma }_{2}}\right) B_{n\text{ \ }%
}\Longrightarrow \ \ B,
\end{equation*}%
and hence by continuous mapping theorem
\begin{equation*}
\left( \frac{\widehat{\sigma }}{\overline{\sigma }_{2}}\right) \sup_{\tau
\in \lbrack 0,1]}\left\vert B_{n}(\tau )\right\vert \underrightarrow{\text{
\ \ \ } \text{ \ \ }}\sup_{\tau \in \lbrack
0,1]}\left\vert B(\tau )\right\vert .
\end{equation*}%
To achieve the proof of (\ref{cv2}) it's sufficient to prove that
\begin{equation}
\widehat{\sigma }\underrightarrow{\text{ \ \ }P\text{ \ }}\overline{\sigma }%
_{2}.  \label{cvp}
\end{equation}%
\textbf{\ }Let $\mathcal{F}_{t}=\sigma -field$ $(\varepsilon
_{1},...,\varepsilon _{t})$ and $\digamma =(\mathcal{F}_{n})$ the
corresponding filtration. Then $N_{n}=\sum_{t=1}^{n}\sigma _{t}\varepsilon
_{t}$ is a square integrable martingale adapted to $\digamma ,$ with
increasing process $\left\langle N_{n}\right\rangle =\sum_{t=1}^{n}\sigma
_{t}^{2}.$

By using (\ref{sig}), $\left\langle N_{n}\right\rangle $ satisfies
\begin{equation*}
\frac{\left\langle N_{n}\right\rangle }{n}\underrightarrow{\text{ \ \ \ }}%
\overline{\sigma }_{2}^{2},
\end{equation*}%
therefore (see Duflo (1997), theorem 1.3.15. )

\begin{equation}
\frac{1}{n}\sum_{t=1}^{n}\sigma _{t}\varepsilon _{t}\underrightarrow{\text{
\ }a.s.\text{\ \ }}0,  \label{l1}
\end{equation}%
where $\underrightarrow{\text{ \ }a.s.\text{\ \ }}$ denotes the almost sure
convergence.

\bigskip

Likewise $M_{n}=\sum_{t=1}^{n}\sigma _{t}^{2}(\varepsilon _{t}^{2}-1)$ is a
square integrable martingale adapted to $\digamma ,$ with increasing process
$\left\langle M_{n}\right\rangle =2\sum_{t=1}^{n}\sigma _{t}^{4}$, hence
theorem 1.3.15. in Duflo (1997) implies that
\begin{equation}
\frac{1}{\left\langle M_{n}\right\rangle }\sum_{t=1}^{n}\sigma
_{t}^{2}(\varepsilon _{t}^{2}-1)\underrightarrow{\text{ \ \ \ }}0\text{
almost surely on }\left\{ \left\langle M_{\infty }\right\rangle =\infty
\right\}  \label{l2}
\end{equation}%
where $\left\langle M_{\infty }\right\rangle =\lim_{n\rightarrow \infty
}\left\langle M_{n}\right\rangle .$ Since
\begin{equation}
\left( \sum_{t=1}^{n}\sigma _{t}^{2}\right) ^{2}\leq n\sum_{t=1}^{n}\sigma
_{t}^{4},  \label{cs}
\end{equation}%
The assumption (\ref{sig}) implies that there exist an universal constants $%
0<K_{1}<K_{2}<\infty $ such that
\begin{equation*}
K_{1}<\frac{1}{n}\sum_{t=1}^{n}\sigma _{t}^{2}<K_{2},
\end{equation*}%
this together with (\ref{cs}) implies that $\left\langle M_{n}\right\rangle
\geq 2nK_{1}^{2}$ which implies that $$\left\{ \left\langle M_{\infty
}\right\rangle =\infty \right\} =\Omega $$

 and hence
\begin{equation}
\frac{1}{\left\langle M_{n}\right\rangle }\sum_{t=1}^{n}\sigma
_{t}^{2}(\varepsilon _{t}^{2}-1)\underrightarrow{\text{ \ \ }a.s.\text{\ }}0,
\label{l21}
\end{equation}%
$\ $Since $(\sigma _{t})$ is a bounded deterministic sequence,\textit{\ }%
then there exists an universal $K>0$ such that $\sigma _{t}^{4}\leq K$ for
all $t\geq 1,$ hence $\left\langle M_{n}\right\rangle \leq nK$ for all $n$,
therefore \textit{\ }
\begin{eqnarray*}
\left\vert \frac{1}{n}\sum_{t=1}^{n}\sigma _{t}^{2}(\varepsilon
_{t}^{2}-1)\right\vert &=&\frac{\left\langle M_{n}\right\rangle }{n}%
\left\vert \frac{1}{\left\langle M_{n}\right\rangle }\sum_{t=1}^{n}\sigma
_{t}^{2}(\varepsilon _{t}^{2}-1)\right\vert \\
&\leq &K\left\vert \frac{1}{\left\langle M_{n}\right\rangle }%
\sum_{t=1}^{n}\sigma _{t}^{2}(\varepsilon _{t}^{2}-1)\right\vert ,
\end{eqnarray*}%
using (\ref{l21}), it follows that
\begin{equation}
\frac{1}{n}\sum_{t=1}^{n}\sigma _{t}^{2}(\varepsilon _{t}^{2}-1)%
\underrightarrow{\text{ \ \ }a.s.\text{\ }}0,  \label{l22}
\end{equation}%
By using (\ref{l1})-(\ref{l22}),
\begin{align}
\frac{1}{n}\sum_{t=1}^{n}y_{t}& =\mu +\frac{1}{n}\sum_{t=1}^{n}\sigma
_{t}\varepsilon _{t}  \label{l3} \\
& \underrightarrow{\text{ \ }a.s.\text{\ \ }}\text{ }\mu  \notag
\end{align}%
and

\begin{align}
\frac{1}{n}\sum_{t=1}^{n}y_{t}^{2}& =\mu ^{2}+2\mu \frac{1}{n}%
\sum_{t=1}^{n}\sigma _{t}\varepsilon _{t}+\frac{1}{n}\sum_{t=1}^{n}\sigma
_{t}^{2}+\frac{1}{n}\sum_{t=1}^{n}\sigma _{t}^{2}(\varepsilon _{t}^{2}-1)
\label{l4} \\
& \underrightarrow{\text{ \ }a.s.\text{\ \ }}\text{ }\mu ^{2}+\overline{%
\sigma }_{2}^{2}  \notag
\end{align}%
Combining (\ref{l3}) and (\ref{l4}) we obtain
\begin{align*}
\widehat{\sigma }^{2}& =\frac{1}{n}\sum_{t=1}^{n}y_{t}^{2}-\left( \frac{1}{n%
}\sum_{t=1}^{n}y_{t}\right) ^{2} \\
& \underrightarrow{\text{ \ }a.s.\text{\ \ }}\text{ }\overline{\sigma }%
_{2}^{2}
\end{align*}%
and hence (\ref{cvp}) follows.

\bigskip \textbf{A2. Proof of Theorem 2}

For all $\tau <\tau _{1}$ we have

\begin{equation}
B_{n}(\tau )=B_{n}^{0}(\tau )+B_{n}^{1}(\tau )  \label{b}
\end{equation}%
where
\begin{equation}
B_{n}^{0}(\tau )=\frac{1}{\sqrt{n}\widehat{\sigma }}\sum_{t=1}^{[n\tau
]}(\sigma _{t}\varepsilon _{t}-\widehat{\mu }_{0}),\widehat{\mu }_{0}=\frac{1%
}{n}\sum_{t=1}^{n}\sigma _{t}\varepsilon _{t}  \label{b0t}
\end{equation}%
\begin{equation*}
B_{n}^{1}(\tau )=\frac{1}{\sqrt{n}\widehat{\sigma }}\left\{ [n\tau ]\mu
_{(1)}-\frac{[n\tau ]}{n}\left( [n\tau _{1}]\mu _{(1)}+(n-[n\tau _{1}]+1)\mu
_{(2)}\right) \right\} .
\end{equation*}%
Straightforward computation leads to

\begin{equation*}
\widehat{\sigma }^{2}\underrightarrow{\text{ \ }a.s.\text{\ \ }}\text{ }%
\sigma _{\ast }^{2}=\overline{\sigma }_{2}^{2}+\tau _{1}(1-\tau _{1})(\mu
_{(1)}-\mu _{(2)})^{2}.
\end{equation*}%
Therefore%
\begin{equation}
B_{n}^{0}(\tau )\underrightarrow{\text{ \ \ \ } \text{
\ \ }}\frac{\overline{\sigma }_{2}}{\sigma _{\ast }}B(\tau )  \label{b0}
\end{equation}%
and
\begin{equation*}
\frac{B_{n}^{1}(\tau )}{\sqrt{n}}\underrightarrow{\text{ \ }a.s.\text{\ \ }}%
\text{ }\frac{\tau (1-\tau _{1})(\mu _{(1)}-\mu _{(2)})}{\sigma _{\ast }}.
\end{equation*}%
Hence

\begin{equation}
B_{n}^{1}(\tau )\underrightarrow{\text{ \ }a.s.\text{\ \ }}\text{ }sgn(\mu
_{(1)}-\mu _{(2)})\infty  \label{b1}
\end{equation}%
where $sgn(x)=1$ if $x>0$ and $-1$ otherwise. Finally (\ref{b}), (\ref{b0})
and (\ref{b1}) imply that
\begin{equation*}
\left\vert B_{n}(\tau )\right\vert \underrightarrow{\text{ \ }P\text{ \ }}%
\text{ }+\infty
\end{equation*}%
and then the desired conclusion (\ref{bh1}) holds.

\bigskip \textbf{A3. Proof of Theorem 3}

\begin{equation}
B_{n}(\tau )=B_{n}^{0}(\tau )+B_{n}^{1}(\tau )  \label{bb}
\end{equation}%
where $B_{n}^{0}(\tau )$ is given by (\ref{b0t}) and

\begin{align*}
B_{n}^{1}(\tau )& =\frac{1}{\sqrt{n}\widehat{\sigma }}\left\{
\sum_{t=1}^{[n\tau ]}\mu _{t}-\frac{[n\tau ]}{n}\sum_{t=1}^{n}\mu
_{t}\right\} \\
& =\frac{(\mu _{(2)}-\mu _{(1)})}{\sqrt{n}\widehat{\sigma }}\left\{
\sum_{t=1}^{[n\tau ]}F(t/n,\tau _{1},\gamma )-\frac{[n\tau ]}{n}%
\sum_{t=1}^{n}F(t/n,\tau _{1},\gamma )\right\} .
\end{align*}%
Straightforward computation leads to

\begin{equation*}
\widehat{\sigma }^{2}\underrightarrow{\text{ \ }a.s.\text{\ \ }}\text{ }%
\sigma _{\ast }^{2}=\overline{\sigma }_{2}^{2}+(\mu _{(2)}-\mu
_{(1)})^{2}\left\{ \int_{0}^{1}F^{2}(x,\tau _{1},\gamma )dx-\left(
\int_{0}^{1}F(x,\tau _{1},\gamma )dx\right) ^{2}\right\} .
\end{equation*}%
Therefore for all $\tau \in (0,1)$

\begin{equation*}
\frac{B_{n}^{1}(\tau )}{\sqrt{n}}\underrightarrow{\text{ \ }a.s.\text{\ \ }}%
\frac{(\mu _{(2)}-\mu _{(1)})}{\sigma _{\ast }}T(\tau ),
\end{equation*}%
where

\begin{equation}
T(\tau )=\int_{0}^{\tau }F(x,\tau _{1},\gamma )dx-\tau
\int_{0}^{1}F(x,\tau _{1},\gamma )dx.  \label{Tt}
\end{equation}%
Moreover, there exists $\tau ^{\ast }\in (0,1)$ such that $T(\tau ^{\ast
})\neq 0$, since if we assume that $T(\tau )=0$ for all $\tau \in (0,1)$
then
\begin{eqnarray*}
\frac{dT(\tau )}{d\tau } &=&F(\tau ,\tau _{1},\gamma )-\int_{0}^{1}F(x,\tau
_{1},\gamma )dx \\
&=&0
\end{eqnarray*}%
for all $\tau \in (0,1)$ which implies that $F(\tau ,\tau _{1},\gamma
)=\int_{0}^{1}F(x,\tau _{1},\gamma )dx=C$ for all $\tau \in (0,1)$ or

\begin{eqnarray*}
\mu _{t} &=&\mu _{(1)}+(\mu _{(2)}-\mu _{(1)})C \\
&=&\mu \text{ for all }t\geq 1
\end{eqnarray*}%
and this contradicts the alternative hypothesis $H_{1}.$

\begin{equation*}
\frac{B_{n}^{1}(\tau ^{\ast })}{\sqrt{n}}\underrightarrow{\text{ \ }a.s.%
\text{\ \ }}\frac{(\mu _{(2)}-\mu _{(1)})}{\sigma _{\ast }}T(\tau ^{\ast })
\end{equation*}%
and $T(\tau ^{\ast })\neq 0$ imply that

\begin{equation*}
\left\vert B_{n}(\tau ^{\ast })\right\vert \underrightarrow{\text{ \ }P\text{
\ }}\text{ }+\infty ,
\end{equation*}%
consequently, the desired conclusion (\ref{bh2}) holds.

\bigskip

\textbf{Remark 2}. For the logistic transition, the function $T(\tau )$ in (%
\ref{Tt}) is given by
\begin{eqnarray*}
T(\tau )&=&\frac{1}{\gamma }\left\{ \tau \log \left[ \left( 1+\exp (\gamma
(\tau _{1}-1)\right) )(1+\exp (\gamma \tau _{1}))\right]  \right\} \\
&-& \frac{1}{\gamma }\left\{  \log \left[ \left(
1+\exp (\gamma (\tau _{1}-\tau )\right) )1 + \exp (\gamma \tau _{1}))\right]
\right\} ,
\end{eqnarray*}

and for the exponential transition%
\begin{equation*}
T(\tau )=\sqrt{\frac{\pi }{4\gamma }}\left\{ (\tau -1) erf(\sqrt{%
\gamma }\tau _{1})+ erf (\sqrt{\gamma }(\tau _{1}-\tau ))-\tau  erf (\sqrt{\gamma }(\tau _{1}-\tau ))\right\} ,
\end{equation*}%
where $erf$  is the Error function given by
\begin{equation*}
erf (x)=\frac{2}{\sqrt{\pi }}\int_{0}^{x}\exp (-t^{2})dt.
\end{equation*}

\end{document}